\documentstyle[epsf]{mn}

\newif\ifAMStwofonts

\ifoldfss
  \ifCUPmtlplainloaded \else
    \NewTextAlphabet{textbfit} {cmbxti10} {}
    \NewTextAlphabet{textbfss} {cmssbx10} {}
    \NewMathAlphabet{mathbfit} {cmbxti10} {} 
    \NewMathAlphabet{mathbfss} {cmssbx10} {} 
  \fi
  \ifAMStwofonts
    \ifCUPmtlplainloaded \else
      \NewSymbolFont{upmath} {eurm10}
      \NewSymbolFont{AMSa} {msam10}
      \NewMathSymbol{\upi}     {0}{upmath}{19}
      \NewMathSymbol{\umu}     {0}{upmath}{16}
      \NewMathSymbol{\upartial}{0}{upmath}{40}
      \NewMathSymbol{\leqslant}{3}{AMSa}{36}
      \NewMathSymbol{\geqslant}{3}{AMSa}{3E}

    \fi
  \fi
\fi 

\ifnfssone
  \newmathalphabet{\mathit}
  \addtoversion{normal}{\mathit}{cmr}{m}{it}
  \addtoversion{bold}{\mathit}{cmr}{bx}{it}
  \newmathalphabet{\mathbfit} 
  \addtoversion{normal}{\mathbfit}{cmr}{bx}{it}
  \addtoversion{bold}{\mathbfit}{cmr}{bx}{it}
  \newmathalphabet{\mathbfss} 
  \addtoversion{normal}{\mathbfss}{cmss}{bx}{n}
  \addtoversion{bold}{\mathbfss}{cmss}{bx}{n}
  \ifAMStwofonts
    \ifCUPmtlplainloaded \else
      \UseAMStwoboldmath
      \makeatletter
      \new@mathgroup\upmath@group
      \define@mathgroup\mv@normal\upmath@group{eur}{m}{n}
      \define@mathgroup\mv@bold\upmath@group{eur}{b}{n}
      \edef\UPM{\hexnumber\upmath@group}
      \new@mathgroup\amsa@group
      \define@mathgroup\mv@normal\amsa@group{msa}{m}{n}
      \define@mathgroup\mv@bold\amsa@group{msa}{m}{n}
      \edef\AMSa{\hexnumber\amsa@group}
      \makeatother
      \mathchardef\upi="0\UPM19
      \mathchardef\umu="0\UPM16
      \mathchardef\upartial="0\UPM40
      \mathchardef\leqslant="3\AMSa36
      \mathchardef\geqslant="3\AMSa3E
    \fi
  \fi
\fi 

\ifnfsstwo
  \DeclareMathAlphabet{\mathbfit}{OT1}{cmr}{bx}{it}
  \SetMathAlphabet\mathbfit{bold}{OT1}{cmr}{bx}{it}
  \DeclareMathAlphabet{\mathbfss}{OT1}{cmss}{bx}{n}
  \SetMathAlphabet\mathbfss{bold}{OT1}{cmss}{bx}{n}
  \ifAMStwofonts
    \ifCUPmtlplainloaded \else
      \DeclareSymbolFont{UPM}{U}{eur}{m}{n}
      \SetSymbolFont{UPM}{bold}{U}{eur}{b}{n}
      \DeclareSymbolFont{AMSa}{U}{msa}{m}{n}
      \DeclareMathSymbol{\upi}{0}{UPM}{"19}
      \DeclareMathSymbol{\umu}{0}{UPM}{"16}
      \DeclareMathSymbol{\upartial}{0}{UPM}{"40}
      \DeclareMathSymbol{\leqslant}{3}{AMSa}{"36}
      \DeclareMathSymbol{\geqslant}{3}{AMSa}{"3E}
    \fi
  \fi
\fi 

\ifCUPmtlplainloaded \else
  \ifAMStwofonts \else 
    \def\upi{\pi}
    \def\umu{\mu}
    \def\upartial{\partial}
  \fi
\fi

\title{Combining the baryon budget with CMBR measurements}
\author[S. Hannestad]
       {S. Hannestad \\
        Institute of Physics and Astronomy,
University of Aarhus,
DK-8000 \AA rhus C, Denmark \\
email: sth@obs.aau.dk}
\date{\today}

\pubyear{1999}

\begin{document}

\maketitle

\label{firstpage}

\begin{abstract}
Measurements of the Cosmic Microwave Background Radiation (CMBR) provide a 
powerful tool for measuring the primary cosmological parameters.
However, there is a large degree of parameter degeneracy in 
simultaneous measurements of the matter density, $\Omega_m$, and 
the Hubble parameter, $H_0$. 
In the present paper we use the presently available CMBR data together
with measurements of the cosmological baryon-to-photon ratio, $\eta$,
from Big Bang nucleosynthesis, and the relative mass fraction of baryons
in clusters to break the parameter degeneracy in measuring 
$\Omega_m$ and $H_0$. 
We find that present data is inconsistent with the standard $\Omega=1$,
matter dominated model. Our analysis favours a medium density universe
with a rather low
 Hubble parameter. This is compatible with new measurements
of type Ia supernovae, and the joint estimate of the two parameters
is $\Omega_m = 0.45^{+0.07}_{-0.07}$ and $H_0 = 39^{+14}_{-13}$
km s$^{-1}$Mpc$^{-1}$. We stress that the upper bound on the Hubble
parameter is likely to be much more uncertain than indicated here, because
of the limited number of free parameters in our analysis.
\end{abstract}



\section{introduction}

Many recent measurements seem to indicate that our universe is {\it not}
a critical density, matter dominated Friedmann universe. Rather,
the recent measurements of type Ia supernovae 
\cite{P98,P97,garnavich,riess,schmidt} strongly suggest
that, although the universe has a flat geometry, the matter density
is low \cite{P98,riess}.
The energy density is instead dominated by either a cosmological
constant \cite{CPT92}, 
or by a similar type of energy with negative pressure,
such as quintessence \cite{wang,zlatev,caldwell}.

Numerous investigations have shown that the fluctuation spectrum of the
cosmic microwave background radiation (CMBR) provides a potentially
very powerful tool for determining cosmological parameters
\cite{jungman,BET97,ZSS97,HET1}.
Thus, one could hope that accurate measurements of these fluctuations
could resolve the issue of whether or not the universe is dominated
by vacuum energy. Based on present observations 
\cite{linbar,QMAP,PYTHON} there already exist
a number of estimates of the cosmological parameters
\cite{efstat1,webster,hancock,linbar,bond,bernardis,tegmark,lineweaver,white}.

There is, however, a severe problem in that a change in one parameter
can often be mimicked by  suitable changes in a combination of other
parameters \cite{HET1,EHT1,EHT2}. 
This is for instance very problematic when trying to
simultaneously determine the cosmological matter density, $\Omega_m$,
and the Hubble parameter, $H_0$.
It has therefore been suggested that CMBR measurements should be 
combined with other constraints, coming for instance from large scale
structure surveys \cite{EHT1,EHT2,webster}
or supernova type Ia measurements
\cite{tegmark,white,lineweaver,EHT1,EHT2,efstat1}.
In this paper we explore another possibility, namely how the cosmic
baryon abundance can be used together with CMBR to provide tight
constraints on $\Omega_m$ and $H_0$.

Galaxy clusters contain large amounts of hot, X-ray emitting gas
from which the baryon mass to total mass ratio can be measured
\cite{WF95,DJF95,EMN96,E97}.
It turns out that measurements of this baryon cluster fraction can
break the parameter degeneracy inherent in the CMBR measurements, and
although most of the present CMBR data is of relatively low accuracy
\cite{linbar,QMAP,PYTHON},
it is still sufficiently accurate to provide good constraints on 
the relevant cosmological models. We have used the presently
available data, together with data from Big Bang nucleosynthesis
and measurements of the cluster baryon fraction,
and find that the observations are indeed strongly incompatible with
a critical density matter dominated model.
Our results are easily compared with for instance the new supernova
type Ia measurements \cite{P98}, 
and are found to be completely compatible.

There still remains a quite large uncertainty on the determination of 
$\Omega_m$ and $H_0$ because of the low accuracy of the CMBR data.
However,
in the very near future, CMBR data of very high quality should become
available from several different experiments. There is the balloon
borne experiment BOOMERANG \cite{boom}
which has already been flown. Also, there
are two new satellite experiments, MAP and PLANCK
\footnote{For information on these missions see the Internet
pages for MAP \\ (http://map.gsfc.nasa.gov) and PLANCK
(http://astro.estec.esa.nl/Planck/).}
which will measure
the fluctuation spectrum very accurately on sub-degree scales.
This should provide data which is accurate enough to diminish the
uncertainties by an order of magnitude.


\section{breaking degeneracy}

As mentioned above, parameter extraction from the CMBR data suffers
from some very large parameter degeneracies \cite{EHT1,EHT2}. 
For instance it is not 
possible to constrain $\Omega_m$ and $H_0$ separately, effectively
only the combination $\Omega_m h^2$ \cite{EHT1,EHT2,HET1}, where
\begin{equation}
h = \frac{H_0}{100 {\rm km}\, {\rm s}^{-1} \, {\rm Mpc}^{-1}},
\end{equation}
can be measured accurately.

However, as shown previously by several authors, it is possible to
break this degeneracy by combining CMBR data with other, complementary,
measurements.
It was shown by Eisenstein {\it et al.} \cite{EHT1,EHT2} 
that large scale structure
surveys like the Sloan Digital Sky Survey (SDSS)
\footnote{See for instance http://www.astro.princeton.edu/BBOOK/}
can probe the 
combination $\Omega_m h^{-1/3}$. This means that the error ellipses
from such surveys are almost orthogonal to those of CMBR. The joint
likelihood function should show a spike-like structure at the crossing
of the two ellipses instead of the very elongated structure of either
of the two individual measurements \cite{EHT1,EHT2}.
In fact, Webster {\it et al.} \cite{webster}
have already performed a joint analysis
of the present CMBR data together with data from the IRAS 1.2 Jy
galaxy survey \cite{IRAS}. 
They find that the CMBR likelihood contours are 
indeed narrowed significantly in the $(\Omega_m,H_0)$ plane.
The above way of breaking degeneracy will work when combining CMBR with any
type of measurement that does not depend on $\Omega_m$ and $H_0$
as $\Omega_m h^2$.

One such possibility is to use the cluster baryon fraction. It has long 
been known that measurements of the baryon cluster fraction favour
a low density universe because the measured fraction is so high that
it cannot support $\Omega_m=1$ without violating BBN constraints
\cite{bludman,hata,white93,E97}.
A standard assumption in this game is to assume that the cluster
baryon fraction is the same as the universal fraction, an assumption
usually referred to as the {\it fair sample hypothesis}.
Numerical simulation seem to justify this assumption. In fact recent
simulations indicate that the cluster baryon fraction is slightly
lower that that of the universe as a whole
(see \cite{hata} for a discussion).
One problem is that observed clusters have diverging baryon fractions,
depending on their total mass. It is argued by Evrard {\it et al.}
\cite{EMN96}
that this is most likely due to errors in the estimate of the total
cluster mass, and that it can be corrected by use of statistical
methods. In this paper we shall use the results obtained by Evrard
\cite{E97} for the universal cluster baryon fraction.

The method for estimating $\Omega_m$ and $H_0$ then works as follows:
Big Bang nucleosynthesis (BBN) can be used to measure the baryon-to-photon
ratio \cite{kolb}, $\eta$, which is related to $\Omega_b$ by
\begin{equation}
\Omega_b h^2 = 3.66 \times 10^7 \, \eta.
\end{equation}
Thus, nucleosynthesis only measures $\Omega_b h^2$, not $\Omega_b$.

On the other hand, the cluster baryon fraction, $f_B$, is really
measured as \cite{bludman,E97}
\begin{equation}
f_B h^{3/2} = \frac{\Omega_b}{\Omega_m} h^{3/2}.
\end{equation}
By combining BBN measurements with the cluster baryon fraction
we can therefore constrain the combination
\begin{equation}
\Omega_m h^{1/2}.
\end{equation}
This combination of $\Omega_m$ and $h$ is sufficiently different from
$\Omega_m h^2$ that when combining them it becomes possible to
constrain both parameters well.


\section{Measurements}

\subsection{CMBR measurements}

In general, the fluctuations in the CMBR is measured in terms of
spherical harmonics
\begin{equation}
T(\theta,\phi) =\sum_{lm}a_{lm}Y_{lm}(\theta,\phi),
\end{equation}
where the coefficients are related to the power-spectrum
$C_l$ coefficients by
\begin{equation}
C_l \equiv \langle|a_{lm}|^2\rangle_m.
\end{equation}
At present there is a host of different CMBR experiments, ranging from
the largest scales (COBE), down to very small scales. The data that 
we use are based on the compilation by Lineweaver and Barbosa
\cite{linbar},
but with the addition of the new Python V \cite{PYTHON}
results, and the results
from the QMAP experiment \cite{QMAP}.


\subsection{BBN measurements}

The past few years have seen a very large fluctuation in the estimated
baryon-to-photon ratio (see for instance \cite{steigman} for a review). 
It has long been known that the primordial
value of deuterium provides a very sensitive probe of $\eta$
\cite{kolb}, but the
problem has been to measure the primordial deuterium abundance. In
the local interstellar medium, the abundance is quite well determined
\cite{hata2},
\begin{equation}
D/H = 1.6 \times 10^{-5},
\end{equation}
but this can only really be used to provide a strict lower limit to
the primordial abundance since deuterium is only destroyed, not 
produced, in astrophysical environments.

Measurements of deuterium in quasar absorption systems at high redshift have
provided a completely new way of measuring the primordial abundance
because such systems are chemically unevolved and should therefore
contain deuterium abundances close to the primordial \cite{steigman}.

There have been two conflicting estimates of the deuterium
abundance in these systems, 
one which is much higher than the local value
\cite{songaila,carswell,rugers,webb,tytler}, 
\begin{equation}
D/H \simeq 2 \times 10^{-4},
\end{equation}
and one which is
only a factor of two higher than the local value
\cite{burles,BT98},
\begin{equation}
D/H \simeq 3 \times 10^{-5}.
\end{equation}

There is growing evidence that the low deuterium value is the correct
one, and as observational values we shall take the so-called 
Low-deuterium/High-helium data set, which is given by
\cite{burles,BT98,IT98}
\begin{eqnarray}
Y_P & = & 0.245 \pm 0.002 \\
D/H & = & (3.4 \pm 0.3) \times 10^{-5}.
\end{eqnarray}
This set of data is completely compatible with standard Big Bang 
nucleosynthesis for a baryon-to-photon ratio of 
\begin{equation}
\eta = (5.1 \pm 0.3) \times 10^{-10},
\end{equation}
or, in terms of baryon density $\Omega_b h^2 = 0.019 \pm 0.001$.


\subsection{Baryon cluster fraction}

As mentioned above, Evrard \cite{E97}
has calculated the universal cluster baryon
fraction based on a large number of clusters. His result is
\begin{equation}
f_B = (0.060 \pm 0.003) h^{-3/2}.
\end{equation}
The $1\sigma$ uncertainty is perhaps somewhat underestimated in the
measurement, and we shall follow Hata {\it et al.} \cite{hata}
in assuming taking the 
$1\sigma$ uncertainty to be twice as large, i.e.\ 0.006.
The above value is derived from the gas fraction alone. The fraction
in hot gas relative to collapsed baryonic objects
has been estimated by White {\it et al.}
\cite{white93} to be
\begin{equation}
\frac{M_{\rm gas}}{M_{\rm gal}} = 5.5 h^{-3/2}
\end{equation}
which is large enough to be insignificant.


\section{Likelihood analysis}

In order to estimate the underlying cosmological parameters, we have 
calculated a large number of synthetic models which are then compared
with the different data sets. 
Our synthetic models range through a large parameter space
and have all been calculated using the publicly available CMBFAST
code \cite{CMBFAST}.
We restrict the discussion to strictly flat models, i.e.\ models with
$\Omega_m + \Omega_\Lambda = 1$. As free parameters we choose
the normalisation, $Q$, the matter density, $\Omega_m$, the baryon
density, $\Omega_b$, the Hubble parameter, $h$, and the spectral
index, $n$. All the above parameters are allowed to vary with the 
restrictions described in table I. Altogether, 65000 independent
CMBR spectra have been calculated, and, apart from the fact that
we do not investigate open models, this analysis is comparable to that
of Lineweaver \cite{lineweaver}. 

In a new analysis, Tegmark \cite{tegmark} has calculated the likelihood
function in a larger, 9-dimensional parameter space. The extra parameters
are: the optical depth to reionisation, $\tau$, and the amplitude
and spectral index of tensor fluctuations.
Here, it was shown that relaxing the assumption about zero curvature
significantly loosens the constraint on the combination $\Omega_m h^2$.
In fact Tegmark finds that there is no upper bound on $\Omega_m h^2$,
even at the 68\% confidence limit, contrary to our findings below.
The reason for this is that the constraint relies to a certain extent
on the amplitude differences in the power spectrum at different $l$.
The information stored in this difference is sensitive to changing
the spatial curvature, and therefore relaxing the flatness assumption
leads to a much looser upper bound on $\Omega_m h^2$.
This being said, there are good reasons for neglecting spatial curvature.
Flatness is a generic prediction of almost all inflationary models, and
furthermore the new data on Supernovae of type Ia clearly favour a
spatially flat universe \cite{P98}.

From the same prejudice that has led us to consider only
flat models, we have neglected possible tensor fluctuations. Almost
all inflationary models predict only very small tensor 
fluctuations, so this is likely to be a good assumption. 
We have also neglected reionisation in our calculations. There is
no justification for this since we know for a fact that the universe 
was reionised at fairly high redshift. As shown by Tegmark
\cite{tegmark}, adding reionisation to the model parameters broadens
the likelihood function, but does not move the maxima.
Therefore we stress that our likelihood estimates should be considered
optimistic and that they will broaden if reionisation is taken into 
account, but that they should still home in on the correct 
central values.

\begin{table}
\caption{The free parameters used in our analysis}
\begin{tabular}{lcc}\hline\hline
Parameter & range & no.\ grid points \\ \hline
$Q_2$ & 5-40 $\mu$K & 200 \\
$\Omega_m$ & 0.2-1 & 17 \\
$\Omega_b h^2$ & 0.002-0.030 & 15 \\
$h$ & 0.30-0.75 & 16 \\
$n$ & 0.7-1.3 & 16 \\ \hline
\end{tabular}
\end{table}

In order to compare with the different data sets we then perform 
a $\chi^2$ analysis. In general, $\chi^2$ is given as
\begin{equation}
\chi^2 = \sum_{i=1}^{N} \frac{1}{\sigma_i^2}
(F_{i,{\rm obs}}-F_{i,{\rm theory}})^2,
\end{equation}
where $F$ for CMBR data is equal to $\Delta T$ for different values
of $l$. For BBN data, $F$ is equal to either the helium or the deuterium
abundance. $\sigma_i$ is the uncertainty in the given data point.

Here it has been assumed that the experimental errors are of Gaussian nature,
which in reality they are not. Therefore the confidence limits that we obtain
are not strict in a mathematical sense, but should rather be seen as 
indicative.

If the measurements are furthermore assumed to be independent then
the likelihood function is then given by $L \propto e^{-\chi^2/2}$, and 
\begin{equation}
L_{\rm joint} = L_{\rm CMBR} L_{\rm BBN} L_{\rm clusters}.
\end{equation}

\begin{figure}
\begin{center}
\epsfysize=10.5truecm\epsfbox{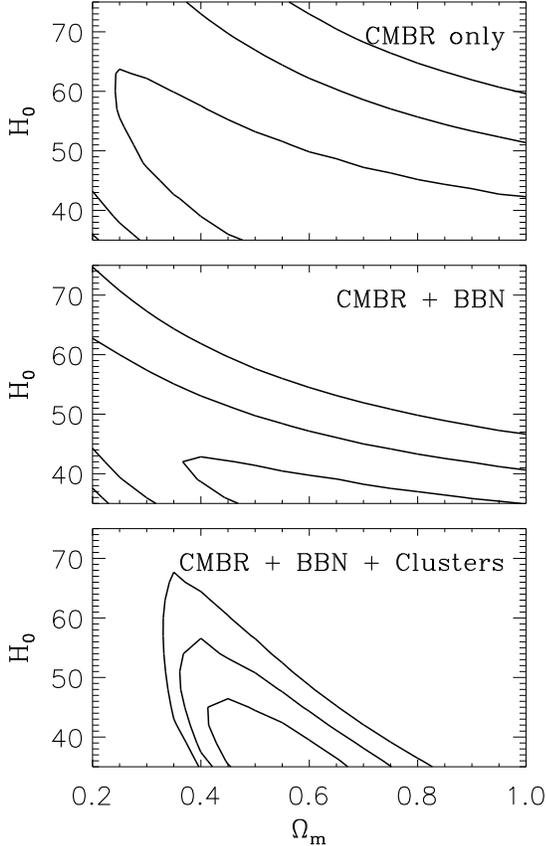}
\vspace*{0.5truecm}
\end{center}
\baselineskip 17pt
\caption{The likelihood function plotted in the ($\Omega_m$,$H_0$)
plane after maximising over all other free parameters.
The contours shown are 68\%, 95\% and 99\% confidence limits.}
\label{fig1}
\end{figure}

In Fig.~1 we show the likelihood contours taking into account different
measurements. 
The likelihood contours have been based on $\Delta \chi^2$, such that
the 68\%, 95\%, and 99\% confidence limits correspond to 
$\Delta \chi^2 = 2.29$, 6.18 and 9.21 respectively.
The top panel shows the likelihood contours for CMBR
data alone.
As expected it shows a very elongated structure, roughly on the
$\Omega_m \propto h^{-2}$ line.
The middle panel shows how the likelihood contours are affected if one
combines CMBR with the BBN estimate of $\Omega_b h^2$. Clearly, there
is only a quite small effect on the determination of $\Omega_m$ and $H_0$,
because $\Omega_b$ is not very degenerate with either of these two parameters.
The real improvement only comes when one takes into account the
cluster baryon fraction measurements. Doing this narrows the likelihood
contours to a more vertical and much tighter structure.
Intriguingly, the combined likelihood function from this method
favours a medium density universe with $\Omega_m$ around 0.5 and a
very low value of the Hubble parameter. This is strongly incompatible
with the standard $\Omega_m=1$ CDM model, unless the Hubble parameter
is much lower than 0.3.
However, as mentioned above our likelihood contours are somewhat optimistic,
especially the upper bound on the Hubble parameter is sensitive to
adding extra parameters to the analysis.

It is also of interest to compare our results with
the recent results from measurements of supernova type Ia's.
The best available data come from the supernova cosmology project,
who have measured 42 high redshift supernovae \cite{P98}. 
From these they obtain
stringent limits on $\Omega_m$ in a flat universe, but no constraint
on the Hubble parameter. Their best fit value for $\Omega_m$ 
in a flat universe is \cite{P98}
\begin{equation}
\Omega_m = 0.28^{+0.09}_{-0.08} ({\rm statistical}) \,
^{+0.05}_{-0.04} ({\rm systematic}).
\end{equation}
These data are compatible with our results, even at the $1\sigma$
level, but only for relatively low values of the Hubble parameter.
Joint maximum likelihood for our analysis and the supernova data
are shown in Fig.~2. The likelihood function has a maximum at
\begin{equation}
(\Omega_m,h) = (0.45,0.39),
\end{equation}
but with a fairly large uncertainty on the Hubble parameter. At the
95\% confidence level the Hubble parameter is only constrained to be
less than 0.64 and, again, we stress that adding more parameters to
the analysis could weaken this bound further.

\begin{figure}
\begin{center}
\epsfysize=6.5truecm\epsfbox{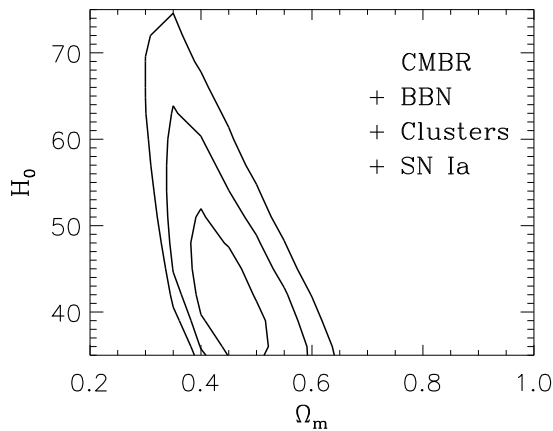}
\vspace{-0.5truecm}
\end{center}
\baselineskip 17pt
\caption{Joint likelihood function from combining the results
of Ref.~\protect\cite{P98} with those of our analysis. The contours
are 68\%, 95\% and 99\% confidence limits.}
\label{fig2}
\end{figure}


\section{Conclusions}

We have calculated the possible constraints on $\Omega_m$ and $H_0$
from combining CMBR data with data on the cosmic baryon abundance.
It was found that these measurements are inconsistent with
the standard flat CDM model, even for very low values of the
Hubble parameter. Our favoured region of parameter space is consistent
with recent measurements from type Ia supernovae at the $1\sigma$
level \cite{P98}.
The joint likelihood function suggests that $\Omega_m$ is close to 0.4
and that a rather low value of the Hubble parameter is favoured.
It is interesting to compare our results with those of Webster 
{\it et al.} \cite{webster}, 
who did a joint analysis of the CMBR data and the IRAS 1.2 Jy
data \cite{IRAS} on large scale structure. They found a best fit for the
joint likelihood of \cite{webster}
\begin{equation}
(\Omega_m,h)_{\rm CMBR+IRAS} = (0.39^{+0.14}_{-0.10},0.53^{+0.05}_{-0.14}),
\end{equation}
which is completely compatible with our estimate of
\begin{equation}
(\Omega_m,h)_{\rm present \, analysis} = 
(0.45^{+0.07}_{-0.07},0.39^{+0.14}_{-0.13}).
\end{equation}

In conclusion what we have shown in the present paper is that if the
cosmic baryons are used in unison with CMBR data to constrain the
cosmological parameters $\Omega_m$ and $h$, the results are
very close to other recent measurements using different methods.
This suggests that
there is a convergence towards the true, underlying,
cosmological model, which is apparently a flat, low density universe.
The fact that so many completely independent methods all yield roughly
consistent estimates make it increasingly unlikely that the data
are contaminated by some large, unknown, systematic error.
Of course when the new high precision data become available
from the BOOMERANG, MAP and PLANCK experiments, 
these uncertainties
should be resolved.

\bsp
\end{document}